\documentclass[final,onefignum,onetabnum]{siamonline190516}
\usepackage{physics}

\usepackage{lipsum}
\usepackage{amsfonts}
\usepackage{epstopdf}
\ifpdf
  \DeclareGraphicsExtensions{.eps,.pdf,.png,.jpg}
\else
  \DeclareGraphicsExtensions{.eps}
\fi

\usepackage{datetime}
\newdateformat{monthyeardate}{%
  \monthname[\THEMONTH], \THEYEAR}

\usepackage{amsmath}
\allowdisplaybreaks
\usepackage{amssymb}
\usepackage{commath}
\usepackage{mathtools}
\usepackage{bbm}

\usepackage{color}
\usepackage{graphicx}
\usepackage[small]{caption}
\usepackage{subcaption}

\usepackage{relsize}
\usepackage{adjustbox}
\usepackage{algorithm}
\usepackage[noend]{algpseudocode}
\usepackage{booktabs}
\usepackage{tikz}

\usepackage{enumitem}
\setlist[enumerate]{leftmargin=.5in}
\setlist[itemize]{leftmargin=.5in}

\newsiamremark{remark}{Remark}
\newsiamremark{hypothesis}{Hypothesis}
\crefname{hypothesis}{Hypothesis}{Hypotheses}
\newsiamthm{claim}{Claim}

\headers{Population heterogeneity in vaccine coverage impacts epidemic thresholds and bifurcation dynamics}{Alina Glaubitz, Feng Fu}

\title{Population heterogeneity in vaccine coverage impacts epidemic thresholds and bifurcation dynamics}

\author{Alina Glaubitz\thanks{Department of Mathematics, Dartmouth College, Hanover, NH 03755, USA (\email{alina.glaubitz.gr@dartmouth.edu}, 
\email{feng.fu@dartmouth.edu})}
\and Feng Fu\footnotemark[1] \thanks{Department of Biomedical Data Science, Geisel School of Medicine at Dartmouth, Lebanon, NH 03756, USA}}

\usepackage{amsopn}

\ifpdf
\hypersetup{
}
\fi

\begin{document}

\maketitle

\begin{abstract}
Population heterogeneity, especially in individuals' contact networks, plays an important role in transmission dynamics of infectious diseases. For vaccine-preventable diseases, outstanding issues like vaccine hesitancy and availability of vaccines further lead to nonuniform coverage among groups, not to mention the efficacy of vaccines and the mixing pattern varying from one group to another. As the ongoing COVID-19 pandemic transitions to endemicity, it is of interest and significance to understand the impact of aforementioned population heterogeneity on the emergence and persistence of epidemics. Here we analyze epidemic thresholds and characterize bifurcation dynamics by accounting for heterogeneity caused by group-dependent characteristics, including vaccination rate and efficacy as well as disease transmissibility. Our analysis shows that increases in the difference in vaccination coverage among groups can render multiple equilibria of disease burden to exist even if the overall basic reproductive ratio is below one (also known as backward bifurcation). The presence of other heterogeneity factors such as differences in vaccine efficacy, transmission, mixing pattern, and group size can each exhibit subtle impacts on bifurcation. We find that heterogeneity in vaccine efficacy can undermine the condition for backward bifurcations whereas homophily tends to aggravate disease endemicity. Our results have practical implications for improving public health efforts by addressing the role of population heterogeneity in the spread and control of diseases.
\end{abstract}

\begin{keywords}
Public health, infectious disease dynamics, bistability, epidemic control
\end{keywords}



\section{Introduction}
\label{intro}
Resurgence of vaccine-preventable diseases, especially the surprising comeback of measles, highlights the importance of addressing pockets of unvaccinated groups or communities with low vaccination rates in order to improve disease control efforts \\\cite{bloom2014addressing}. While the phrase ``the pandemics of unvaccinated'' has been dominating the civil discourse of the current COVID-19 mass vaccination campaign~\cite{bor2022discriminatory}, it is imperative for public health stakeholders to understand the impact of population heterogeneity in various forms on the spread and control of diseases. Doing so will be key to the success of one health initiative given the increasingly connected nature of both local and global populations~\cite{rubin2013review}.

Concerning the ongoing COVID-19 pandemic~\cite{vespignani2020modelling} as well as other epidemics \cite{royce2020mathematically,glaubitz2020oscillatory,teslya2022}, prior studies have considered heterogeneity in various aspects, including individuals' network of contacts~\cite{may2001infection} and vaccine uptake \cite{wang2016statistical}. Furthermore, different vaccines appear to have different efficacies \cite{vc-eff} and even for people taking the same vaccine, different levels of anti-bodies after vaccination suggest heterogeneity in vaccine efficacies between individuals and in particular between different age groups \cite{het-eff1,het-eff2,chen2019imperfect}.

In the context of nonmedical exemptions to school immunization requirements, mask-wearing behavior during the COVID-19 pandemic as well as other COVID-19 measures  \\ \cite{hom-covid,vc-rate-het,hom-risk-cov}, it appears that homophily, i.\,e.\ the tendency for people to seek out or be attracted to those who are similar and share similar opinions \cite{mcpherson2001birds,fu2012evolution}, drives social interactions within and between groups and thus influences the spread of infectious diseases \cite{salathe2008effect,hom-het1,hom-het2,vc-rates,vc-rate-het}. 

Ever since \cite{sir-model} first introduced the Susceptible-Infected-Recovered (SIR) model,  compartmental models of this kind have been used as a quantitative means to understand the spread of infectious diseases in populations. One defining characteristic of these models is the basic reproductive ratio ($R_0$) \cite{R0}. $R_0$ is an epidemic threshold parameter that determines whether one index case or a few initial infected individuals can seed the affected population and cause a disease outbreak. In particular, $R_0$ measures how many secondary infections arise as the result of the introduction of one infectious individual in a totally susceptible population. Relating this parameter $R_0$ to common epidemic models that have a constant influx of susceptibles, for $R_0<1$ there exists one stable disease-free equilibrium. For $R_0=1$ a bifurcation occurs and for $R_0>1$ there exists the unstable disease-free equilibrium as well as a globally stable endemic equilibrium \cite{fb-r0,Li2001}. This epidemic threshold means not only that an outbreak occurs and the disease becomes endemic when $R_0>1$, but also that to eradicate an endemic disease from a population, we need $R_0<1$. This dynamical behavior is called forward bifurcation.

One assumption made in commonly used SIR models is that the population is homogeneous (i.e, well-mixing). However, as discussed before, this is not necessarily the case for real-world scenarios. When heterogeneity is introduced in compartmental models, $R_0$ becomes a weighted average of the spreading capacities across different groups in the population. As a consequence, this can change the predictive character of $R_0$. While the effective $R_0 > 1$ can still tell us whether the disease can invade and cause an outbreak, it may be insufficient for the condition $R_0<1$ to eradicate an endemic disease from the population. This phenomenon is called backward bifurcation, in which one stable and one unstable endemic equilibrium co-exist for $R_0<1$. A variety of mechanisms have been found that introduce backward bifurcation since its first discovery. They include imperfect vaccines and education about prevention over limited treatment availability, non-linear incidence rate, non-constant contact rates, partial protection against reinfection, and varying transmission rates to imperfect lockdowns \cite{bb-covid,HADELER1995,HADELER1997,Brauer2004,KRIBSZALETA2000,HUI2005,KRIBSZALETA2002,NUDEE2019,GARBA2008,bb-ex,VANDENDRIESSCHE2002,GUMEL2012,Zhang2008,bb-treatment,bb-convex,bb-nonlinear,bb-sis,bb-cattle,bb-tb1,bb-first?}. Backward bifurcation has also been found in models for HIV/Aids, Malaria, combination models for HIV/Aids and Malaria, H1N1, Dengue, tuberculosis, and COVID-19 \cite{bb-tb1,bb-first?,huang1992,bb-malaria,bb-hiv-malaria,bb-ex,GARBA2008,asatryan2015new}.

Built on these prior results, our present study aims to understand the combined effect that population heterogeneity and homophily have on the spread and control of infectious diseases. In particular, we consider imperfect vaccines using a SIRV model that has multiple groups. We introduce and also vary the level of, heterogeneity between groups and homophily within groups. Similar to the one-group scenario discussed by \cite{bifurcation_arino}, we show that imperfect vaccination is responsible for causing backward bifurcations (by finding endemic equilibria). Importantly, we find that population heterogeneity in the vaccine coverage can greatly induce backward bifurcations, i.\,e.\ it can cause the emergence of endemic equilibria for $R_0<1$ under a wide range of model parameters. Further subtleties in heterogeneity in vaccine efficacy and susceptibility can further impact conditions for backward bifurcations. Moreover, we show how homophily increases the basic reproductive ratio and can cause the emergence of endemic equilibria in a population that is not well-mixing.

The rest of this paper is organized as follows. Section (\ref{sec:model}) introduces the specific SIRV model that we are studying by accounting for group-dependent heterogeneity of our concern. Section (\ref{sec:results}) presents our analytical findings. In Subsection (\ref{subs:bb}) we derive and obtain the equilibria of the model as the roots of a polynomial of degree 4, and in Subsection (\ref{subs:het}) we discuss how heterogeneity in vaccination rate, transmission rate, and vaccine efficacy influence the model's epidemic thresholds and bifurcation dynamics. Then, in Subsection (\ref{subs:hom}) we present numerical results regarding the effect of homophily. Finally, in Section (\ref{sec:disc}) we discuss our results and the potential implications of our findings.

\section{Model}
\label{sec:model}

Without loss of generality, we use a compartmental model where the population is divided into two groups $\{1,2\}$ with relative sizes $P_1$ and $P_2$ respectively such that $P_1+P_2=1$. Extending this model to multiple (more than two) groups is straightforward, and we confirm qualitatively similar results.  Within each group, we assume well-mixing. Each individual is either susceptible and unvaccinated ($S$), vaccinated ($V$), infected ($I$), or removed ($R$).
The transmission rate is $\beta_i, i \in \{1,2\}$ for (unvaccinated) susceptible individuals in encounters with infected, while vaccinated from group $i$ get infected with rate $\beta_i(1-\varepsilon_i), j \in \{1,2\}$. Infected are removed after being infected for time $1/\gamma$. After time $1/\nu$, people from the removed compartment become susceptible again. After time $1/\psi$, people lose their partial immunity gained through vaccination and move from the vaccinated to the susceptible compartment. Further, people in the vaccinated compartment get vaccinated with rate $\phi_i, i \in \{1,2\}$. 
In particular, we introduce heterogeneity between the two groups in the following ways:
\begin{itemize}
    \item[(i)] Vaccination rates $\phi_i, i \in \{1,2\}$. People get vaccinated faster in one group than in the other (leading to higher coverage in equilibrium).
    \item[(ii)] Vaccine efficacies $\varepsilon_i, i \in \{1,2\}$. The vaccine has different efficacies in each of the groups. 
    \item[(iii)] Susceptibility rates $\beta_i, i \in \{1,2\}.$ Individuals from group $i$ are infected with the disease with rate $\beta_i$ when unvaccinated, while they are infected with rate $\beta_i(1-\varepsilon_i)$ when vaccinated.
\end{itemize}

The mixing pattern for contact rates within each population group is given by $C_{1,1} = C_{2,2}$ and between the groups by $P_2 C_{1,2} = P_1 C_{2,1}$, where $C_{1,1} \geq C_{2,1}$.
Accordingly, the infectious disease dynamics are given by
\begin{align}
    \dot{S_i} &= \mu P_i +\nu R_i - S_i \beta_i \left(C_{i,1} I_1 + C_{i,2} I_2\right) -(\phi_i+\mu) S_i + \psi V_i\\ \nonumber
    \dot{V_i} &= -V_i (1-\varepsilon_i) \beta_i \left(C_{i,1} I_1 + C_{i,2} I_2\right)  + \phi_i S_i - (\psi+\mu) V_i\label{eq:model}\\
    \dot{I_i} &= ( S_i + (1-\varepsilon_i) V_i)  \beta_i \left(C_{i,1} I_1 + C_{i,2} I_2\right)  - (\gamma+\mu) I_i\nonumber\\
    \dot{R_i} &= \gamma I_i -(\nu+\mu) R_i \nonumber
\end{align}
for $i \in \{1,2\}$, $i \neq j$ with initial condition $S_i(0) = S_i^0, V_i(0) = V_i^0, I_i(0) = I_i^{0}, R_i(0) = P_i-(S_i^0+V_i^0+I_i^0)$.

\begin{figure}[tb]
    \centering
    \includegraphics[width=\linewidth]{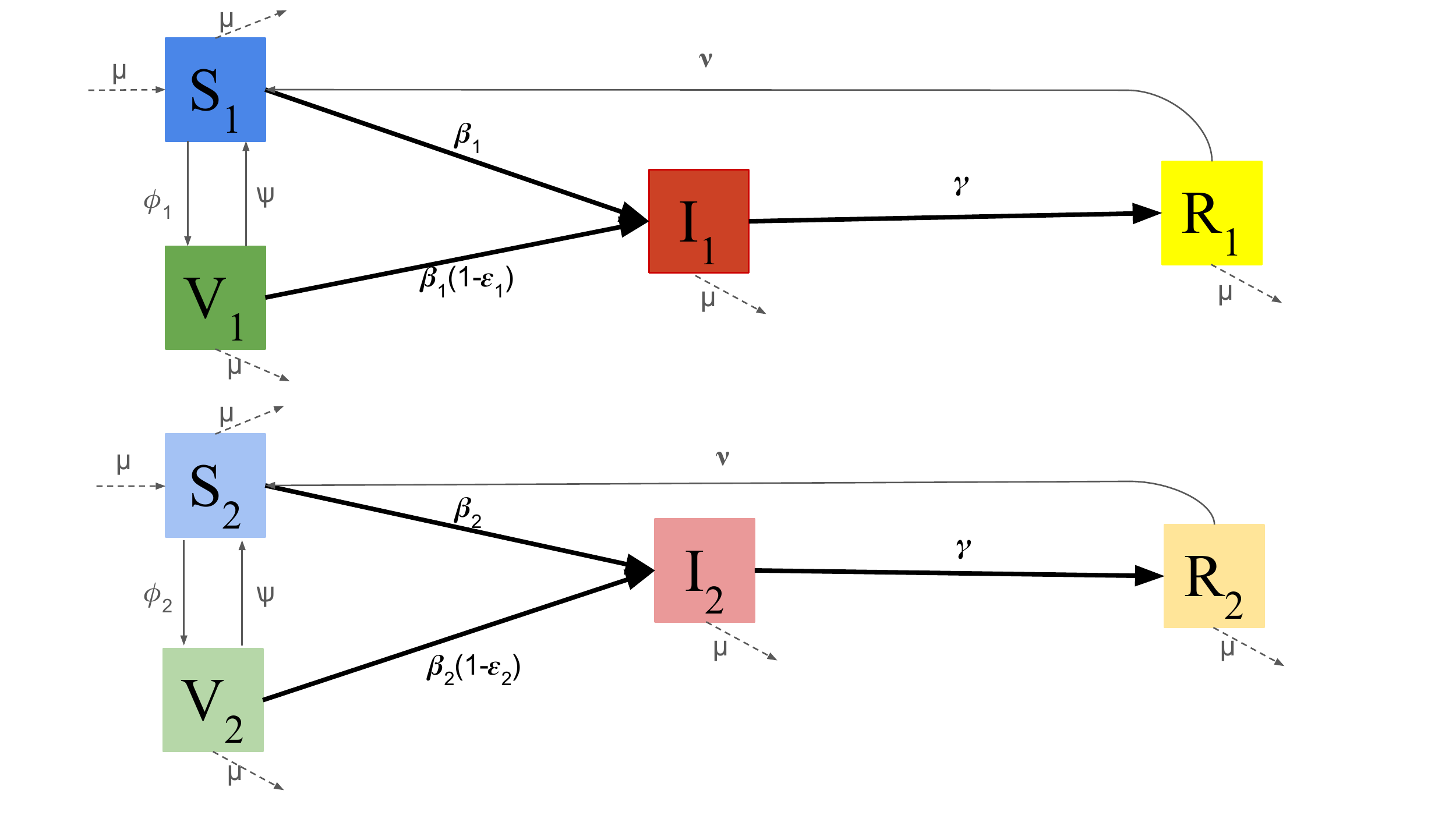}
    \caption{Schematic of the model. Here, we consider two groups to account for heterogeneity between the groups in the vaccine coverage, the vaccine efficacy, and the susceptibility to infection.}
    \label{fig:modified}
\end{figure}

 A schematic of this model is given in Figure \ref{fig:modified}.

In what follows, we show that the model above exhibits backward bifurcations under certain conditions and we are able to find an explicit formula for its equilibria for simpler cases. We find that heterogeneity and homophily can cause a qualitative change in the model's dynamical behavior. In particular, heterogeneity in vaccine coverage can cause backward bifurcations when a corresponding homogeneous population does not exhibit backward bifurcations at all.

Heterogeneity in vaccine efficacy and susceptibility can also impact the conditions for backward bifurcations, and in the presence of vaccine coverage heterogeneity, subtleties of heterogeneity in vaccine efficacy and susceptibility can further change these conditions. Moreover, homophily increases the basic reproductive ratio $R_0$ and can cause a disease to become endemic.

We detail our analysis and present comparative results regarding these findings as below.

\section{Results}
\label{sec:results}

\subsection{Conditions for backward bifurcation} \label{subs:bb}

We computationally analyze the bifurcation dynamics of the full model with respect to varying model parameters and investigate their impact on the conditions for backward bifurcation. To obtain analytical intuitions, here we make a simplifying assumption: the population has two groups of equal size and is fully well-mixing, i.\,e.\ $C_{i,j}=1$ for $i,j = 1,2$, which serves as a base case for comparisons with our general results.

We compute the basic reproductive ratio in this scenario via the next-generation method as described by \cite{R0}. Following the notation of this paper, we define the rate with which new infections arise in each compartment $F_j, j \in \{1,2\},$ and the rate with which infections are transferred between compartments as $V_j, j \in \{1,2\}$. Further, we define
\[
    F = \left[ \frac{\partial F_i}{\partial I_j}\right], \qquad V = \left[ \frac{\partial V_i}{\partial I_j}\right].
\]
Then, the basic reproductive ratio is given by
\[
    R_0 = \max \lambda(FV^{-1}) = \text{largest eigenvalue of }FV^{-1}.
\]
In particular, we obtain 
\begin{align*}
    R_0 &= \frac{1}{\gamma+\mu}\left( \beta_1 P_1 \frac{\psi+\mu+(1-\varepsilon_1)\phi_1}{\psi+\mu+\phi_1} + \beta_2 P_2 \frac{\psi+\mu+(1-\varepsilon_2)\phi_2}{\psi+\mu+\phi_2}  \right).
\end{align*}

The equilibria are given by the solutions of 
\begin{equation} \label{equilibrium}
        \dot{S_i} = \dot{V_i} = \dot{I_i} =\dot{R_i} = 0,\quad i = 1,2.
\end{equation}
The disease-free equilibrium (DFE)
\begin{equation} \label{DFE}
    (S_1^{\text{DFE}},V_1^{\text{DFE}},0,0,S_2^{\text{DFE}},V_2^{\text{DFE}},0,0),
\end{equation}
where
\begin{align*}
    S_i^{\text{DFE}} = \frac{\psi + \mu}{\psi + \phi_i + \mu}P_i, \qquad
    V_i^{\text{DFE}} = \frac{\phi_i}{\psi + \phi_i + \mu}P_i, \qquad
\end{align*}
exists for any parameters $\psi,\phi_1,\phi_2,\varepsilon_1,\varepsilon_2> 0$. Moreover, we might get an endemic equilibrium for $(\beta_1+\beta_2)/2>\gamma + \mu$, which means that in a population without vaccination the disease becomes endemic. We find a solution for these equilibria by solving \eqref{equilibrium} for $I = I_1+ I_2$.
In the equilibrium, $(\nu+\mu) R_i = \gamma I_i$ as well as $V_i = P_i - S_i - \left(1+\frac{\gamma}{\nu+\mu}\right)I_i$. This then implies that
\begin{align*}
    S_i = \frac{(P_i-(1+\frac{\gamma}{\nu+\mu})I_i)((1-\varepsilon_i)I+\psi+\mu)}{(1-\varepsilon_i)I+\phi_i+\psi+\mu}.
\end{align*}
Hence,
\[
    I_i = \frac{\beta_i(\varepsilon_i P_i(\beta_i(1-\varepsilon_i)I+\psi+\mu)+(1-\varepsilon_i)P_i(\phi_i+\beta_i(1-\varepsilon_i)I+\psi+\mu))}{(\beta_i(1-\varepsilon_i)I+\psi+\mu)(\gamma+\mu+(1+\frac{\gamma}{\nu+\mu})\beta_i I) + \phi_i (\gamma+\mu+(1+\frac{\gamma}{\nu+\mu})(1-\varepsilon_i)\beta_i I)}
\]
This implies
\begin{align*}
    I &= I_1 + I_2 \\
    &= \frac{\beta_1(\varepsilon_1 P_1(\beta_1(1-\varepsilon_1)I+\psi+\mu)+(1-\varepsilon_1)P_1(\phi_1+\beta_1(1-\varepsilon_1)I+\psi+\mu))}{(\beta_1(1-\varepsilon_1)I+\psi+\mu)(\gamma+\mu+(1+\frac{\gamma}{\nu+\mu})\beta_1 I) + \phi_1 (\gamma+\mu+(1+\frac{\gamma}{\nu+\mu})(1-\varepsilon_1)\beta_1 I)} \\
    &\quad + \frac{\beta_2(\varepsilon_2 P_2(\beta_2(1-\varepsilon_2)I+\psi+\mu)+(1-\varepsilon_2)P_2(\phi_2+\beta_2(1-\varepsilon_2)I+\psi+\mu))}{(\beta_2(1-\varepsilon_2)I+\psi+\mu)(\gamma+\mu+(1+\frac{\gamma}{\nu+\mu})\beta_2 I) + \phi_2 (\gamma+\mu+(1+\frac{\gamma}{\nu+\mu})(1-\varepsilon_2)\beta_2 I)}.
\end{align*}
This equation only depends on one variable: $I$, and its solutions are given by $I=0$ as well as the roots of the polynomial $P(I)$
\begin{equation} \label{eq:polynomial}
    0 = P(I) = aI^4+b I^3 + cI^2 + dI +e,
\end{equation}
where $a,b,c,d,e$ depend on the parameters of the model.
\cite{quartic} provides us with exact solutions for the force of infection $I$ and we can then solve for $I_1$ and $I_2$. Note here that our approach can be used to find the equilibria in a model with any number of groups as the roots of a polynomial of degree that is twice the number of groups.
\smallskip

\begin{figure}[tb]
    \centering
    \includegraphics[width=\linewidth]{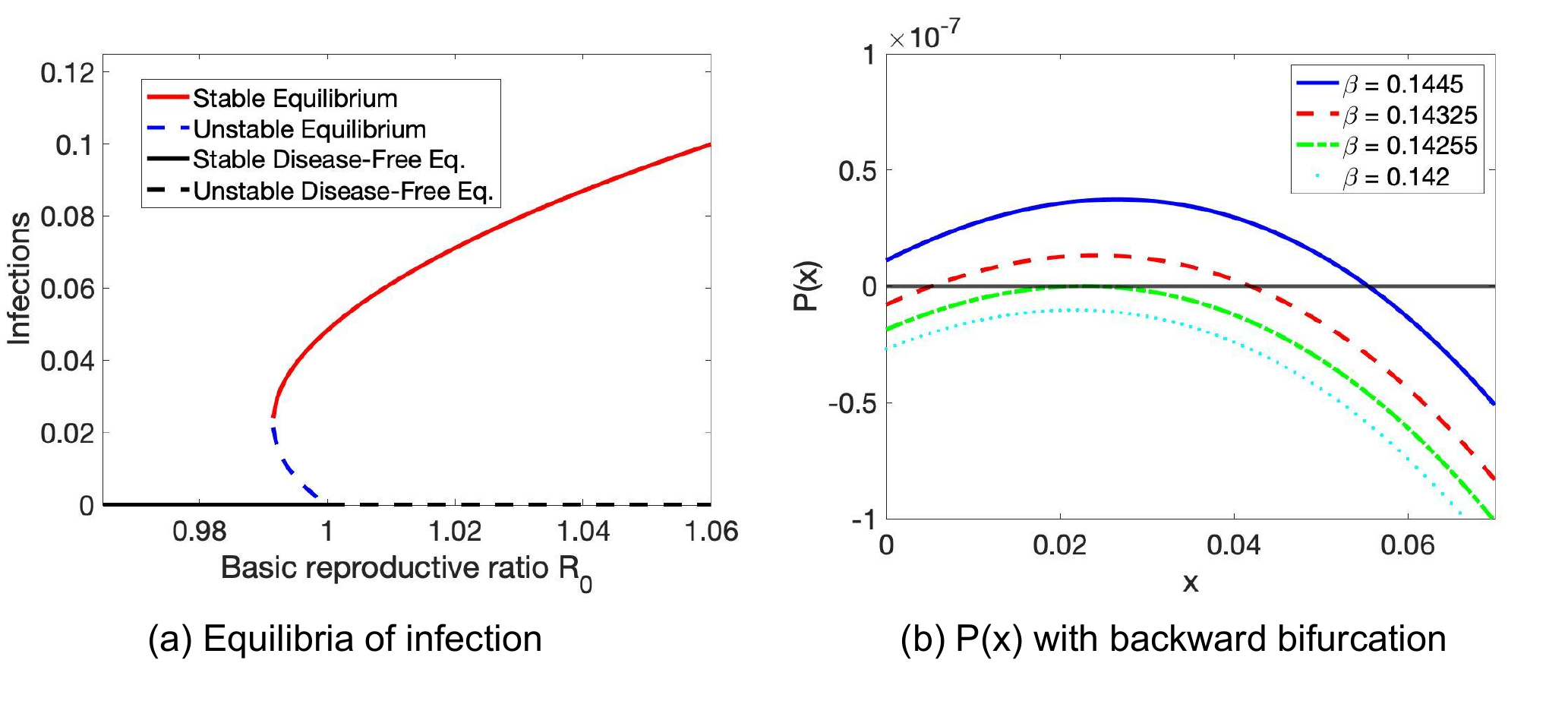}
    \caption{Backward bifurcation in the presence of an imperfect vaccine and population heterogeneity in vaccine coverage.  The Mathematica method Solve was used to solve the polynomial $P(I)$ in Eq.~\eqref{eq:polynomial}.}
    \label{fig:PI}
\end{figure}

Since \eqref{eq:polynomial} is a polynomial of degree 4, it has four roots that are either real or complex. We note that this might lead to up to four solutions. However, we could only find up to two reasonable solutions. For small $R_0\ll1$, the roots are all complex, and the discriminant
 \begin{align*}
 \Delta &=256a^{3}e^{3}-192a^{2}bde^{2}-128a^{2}c^{2}e^{2}+144a^{2}cd^{2}e-27a^{2}d^{4}+144ab^{2}ce^{2}\\&-6ab^{2}d^{2}e-80abc^{2}de+18abcd^{3}+16ac^{4}e\-4ac^{3}d^{2}-27b^{4}e^{2}+18b^{3}cde\\&-4b^{3}d^{3}-4b^{2}c^{3}e+b^{2}c^{2}d^{2}
 \end{align*}
is positive, and we have two complex conjugate solutions. As $R_0$ increases, a bifurcation occurs, at the point where the discriminant
is zero, and one complex conjugate root becomes a double real root. This real root might either be positive or negative. If it is negative, we see the classical forward bifurcation. In this case, $R_0<1$ implies no endemic equilibrium, $R_0>1$ means that there is one endemic equilibrium. However, if the double root is positive, we can instead observe a backward bifurcation. Then, we have two endemic equilibria for $R_{BB}<R_0<1$ for some $1>R_{BB}>0$ where $R_{BB}$ is the lower critical threshold of basic reproductive ratio above which the system exhibits backward bifurcation. The larger one of these two roots is stable while the smaller one is unstable. This behavior is similar to the one-group scenario as described by \cite{bifurcation_arino}. An illustration of bifurcation dynamics can be found in Figure \ref{fig:PI}. Moreover, this behavior is similar to the two-group SIS model from \cite{KRIBSZALETA2000} where the stability properties of the model have been investigated and the existence of an even number of equilibria was proven.

\subsection{Population heterogeneity and backward bifurcation} \label{subs:het}

\begin{figure}[tb]
    \centering
    \includegraphics[width=\linewidth]{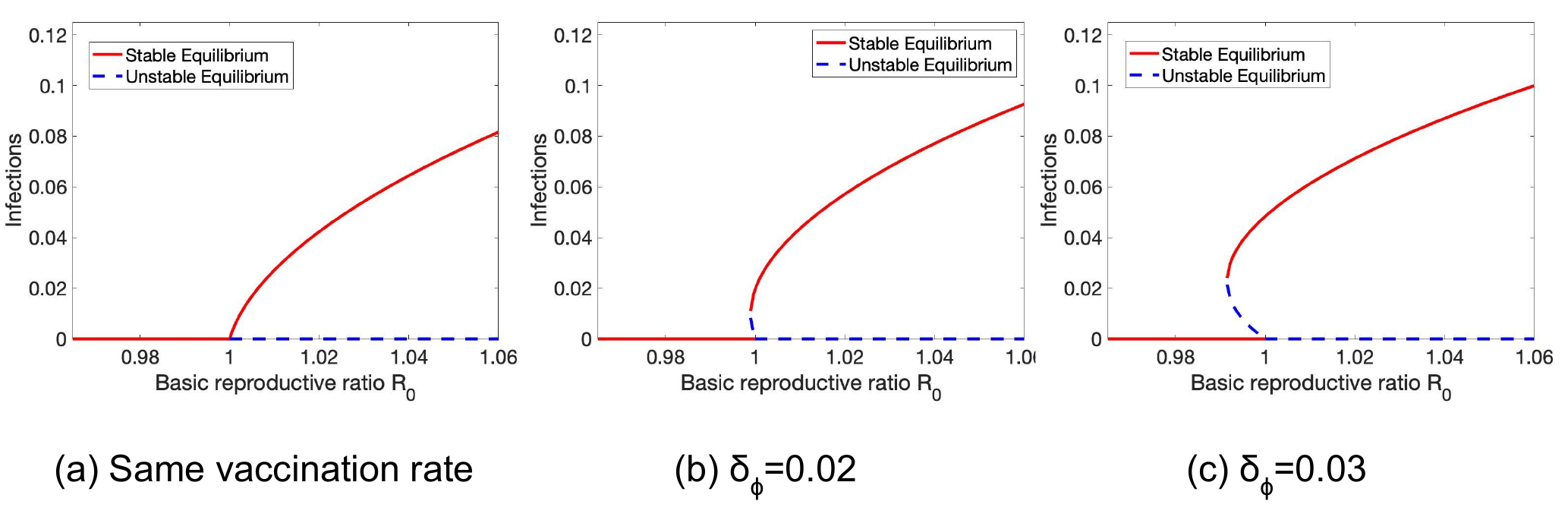}
    \caption{Heterogeneity in vaccine coverage causes backward bifurcations. (a) When the two groups have the same vaccination rate, no backward bifurcation occurs. (b), (c) As the groups become increasingly heterogeneous, we can see the emergence of backward bifurcations. Parameters here are $\gamma = 0.05,\psi = 0.002,\phi = 0.04, \nu = 1/30,\varepsilon=0.725$ and $\beta$ is varying.}
    \label{fig:BB_cov}
\end{figure}

Let us turn to understand the effect of population heterogeneity on the proportion of infected people at equilibrium. Depending on the type of, as well as the level of, population heterogneity, we observe different behaviors. 
Most interestingly, heterogeneity in vaccine coverage (differences in vaccine uptake rates) can cause the emergence of endemic equilibria (see Figure \ref{fig:BB_cov}). Here, we see that for increasing values of heterogeneity $\delta_\phi$ (defined as the absolute difference between the two groups) two endemic equilibria emerge, while no endemic equilibria exist in a homogeneous population.

\begin{figure}[tb]
    \centering
    \includegraphics[width=\linewidth]{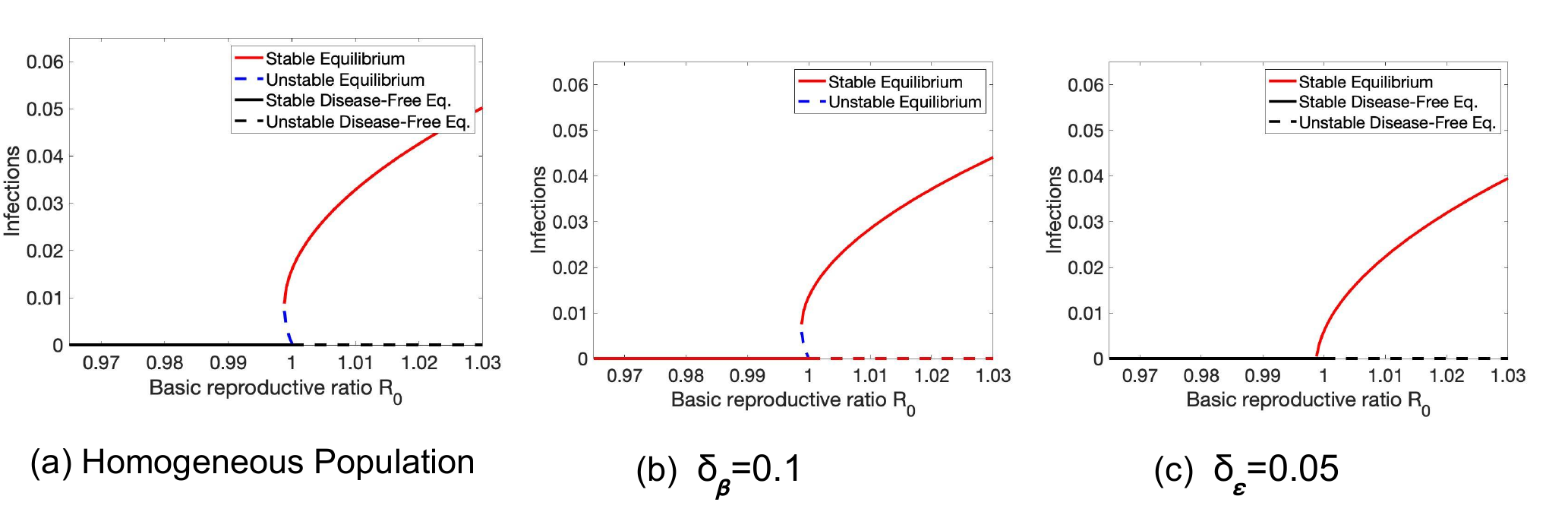}
    \caption{Heterogeneity in vaccine efficacy and susceptibility can impact the conditions for backward bifurcations. Parameters here are $\gamma = 0.05,\varepsilon = 0.85,\nu = 1/60,\psi = 0.001,\phi = 0.05$.}
    \label{fig:BB_eff}
\end{figure}

Further, in this model, heterogeneity in vaccine efficacy and susceptibility can impact the conditions for backward bifurcations. A demonstration of their impacts can be found in Figure \ref{fig:BB_eff}. Here, we see that on their own, both heterogeneity in susceptibility $\delta_\beta$ (transmission rate) and vaccine efficacy $\delta_\varepsilon$ can prevent the existence of endemic equilibria.

\begin{figure}[tb]
    \centering
    \includegraphics[width=\linewidth]{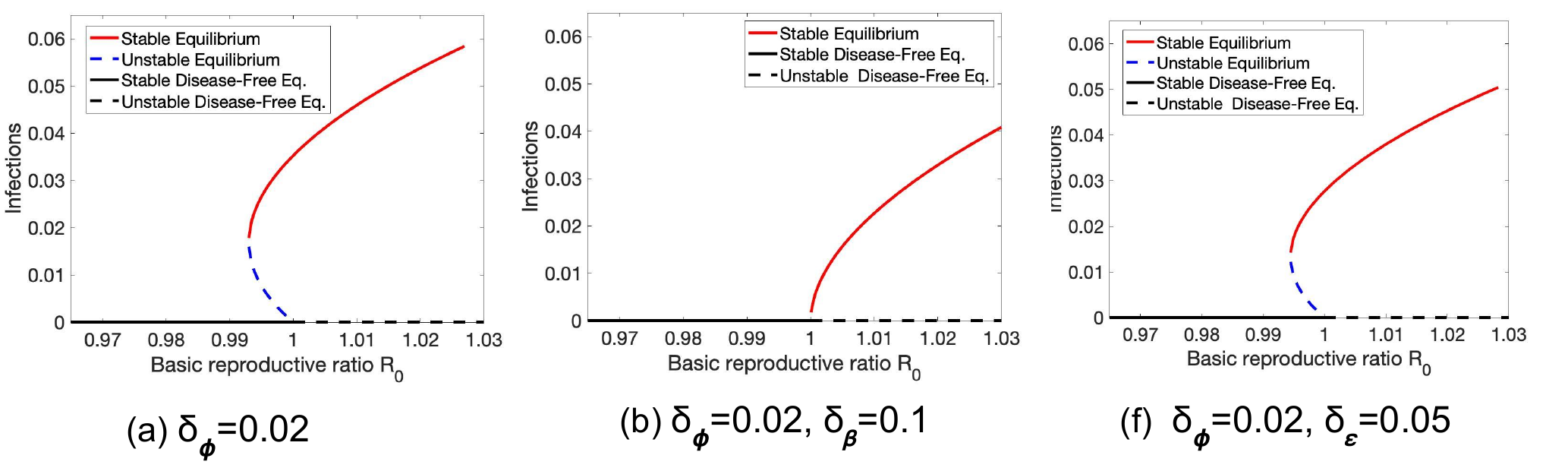}
    \caption{In the presence of vaccine coverage heterogeneity, subtleties of heterogeneity in vaccine efficacy and susceptibility can impact the conditions for backward bifurcations.  Here, both parameters are elevated in one group. Parameters here are $\gamma = 0.05,\varepsilon = 0.85,\nu = 1/60,\psi = 0.001,\phi = 0.05$.}
    \label{fig:BB_eff_cov1}
\end{figure}

\begin{figure}[tb]
    \centering
    \includegraphics[width=\linewidth]{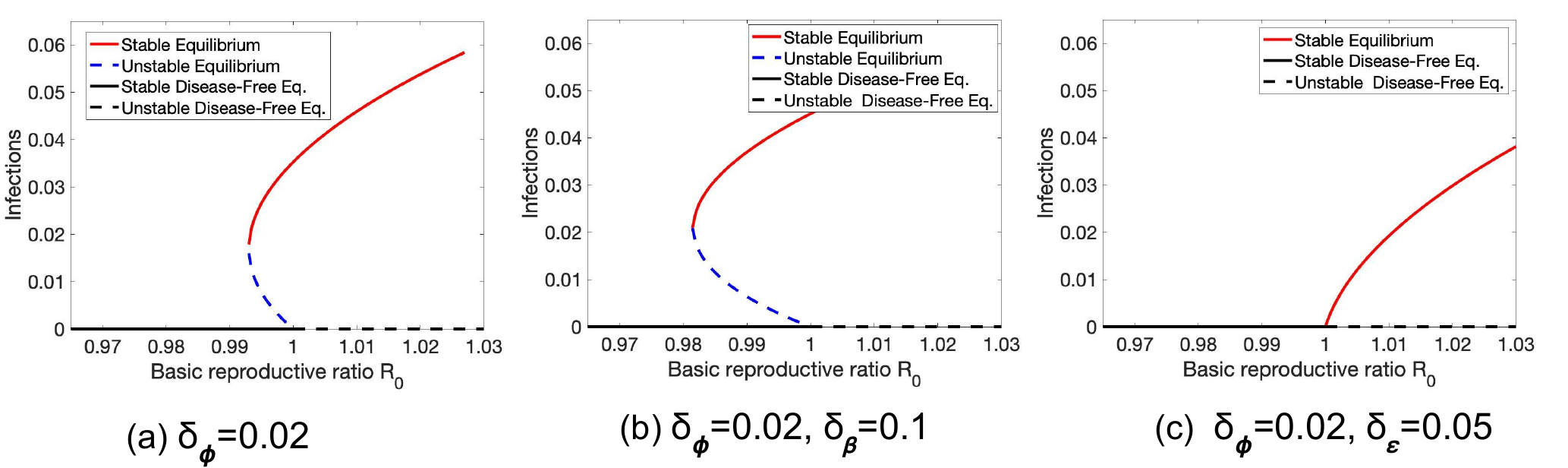}
    \caption{In the presence of vaccine coverage heterogeneity, subtleties of heterogeneity in vaccine efficacy and susceptibility can impact the conditions for backward bifurcations. Here, only one parameter is elevated in each group. Parameters here are $\gamma = 0.05,\varepsilon = 0.85,\nu = 1/60,\psi = 0.001,\phi = 0.05$.}
    \label{fig:BB_eff_cov2}
\end{figure}

However, if we introduce heterogeneity not only in vaccine efficacy or susceptibility but also in vaccine coverage simultaneously, subtleties in their exact compositions can drastically impact the model's behavior. We can see an illustration of their subtle impacts in Figure \ref{fig:BB_eff_cov1} and \ref{fig:BB_eff_cov2}, respectively. In Figure \ref{fig:BB_eff_cov1}, both parameters are elevated in the same group. Here, vaccine coverage heterogeneity alone destabilizes the system Figure \ref{fig:BB_eff_cov1}(a). However, introducing further heterogeneity in susceptibility Figure \ref{fig:BB_eff_cov1} (b) and vaccine efficacy Figure \ref{fig:BB_eff_cov1} (c), stabilizes the system. However, if we elevate one parameter in each group (Figure \ref{fig:BB_eff_cov2}), heterogeneity in susceptibility destabilizes the system Figure \ref{fig:BB_eff_cov2}(b), while heterogeneity in vaccine efficacy stabilizes the behavior Figure \ref{fig:BB_eff_cov2} (c).

\begin{figure}[tb]
    \centering
    \includegraphics[width=.8\linewidth]{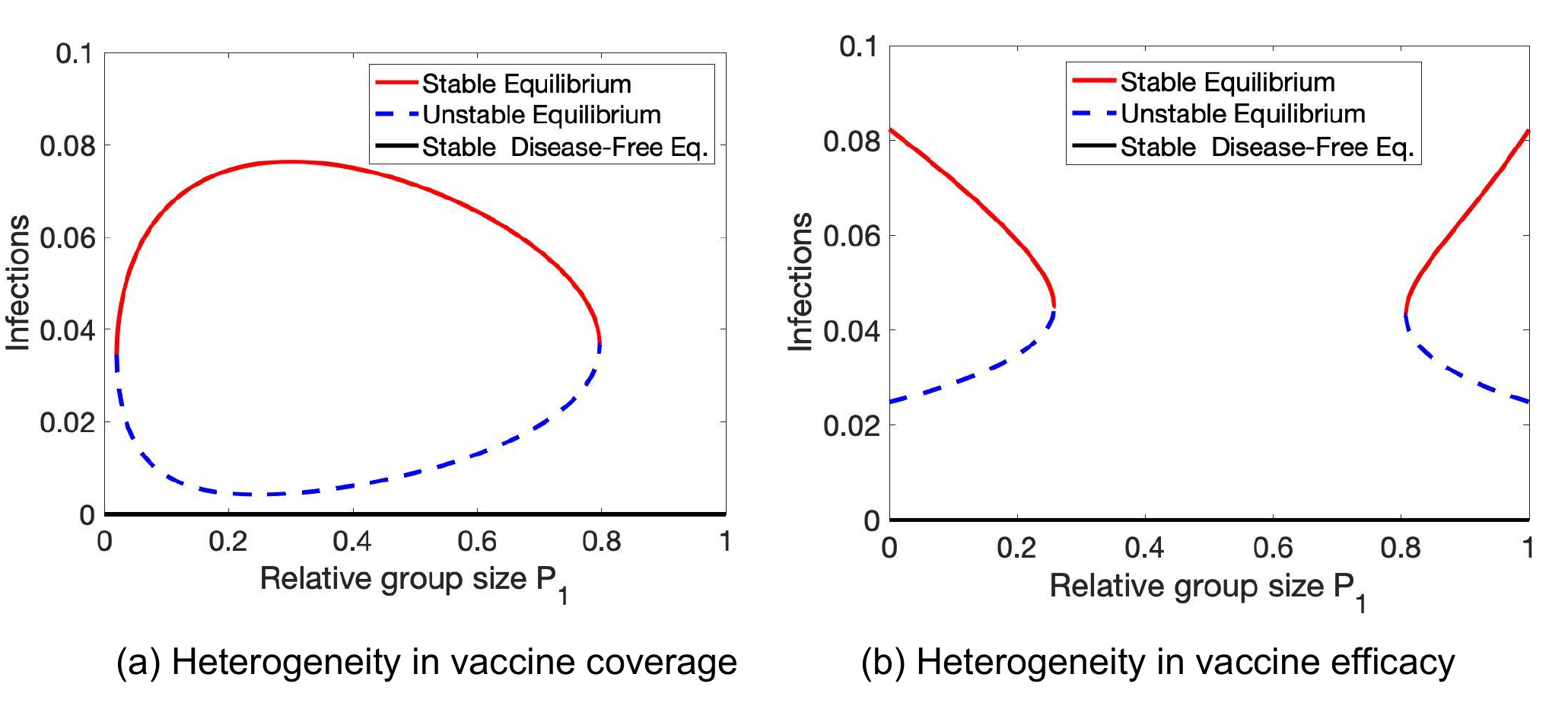}
    \caption{Relative group size matters. The stabilizing and destabilizing effects of vaccine coverage and vaccine efficacy heterogeneity are critically dependent on the composition of the two groups. Parameters are (a) $\beta = 0.5,\gamma = 0.05,\nu=1/30,\psi=1/(5*365),\phi_1 = 0.1,\phi_2 = 0.2,\varepsilon = 0.9$, (b) $\beta = 0.6,\gamma = 0.05,\nu=1/30,\psi=1/(5*365),\phi = 0.15,\varepsilon_1 = 0.65, \varepsilon_2 = 0.95$, respectively (c) $\beta_1 = 0.4,\beta_2 = 0.6,\gamma = 0.05,\nu=1/30,\psi=1/(5*365),\phi = 0.15,\varepsilon = 0.9$.}
    \label{fig:BB_size}
\end{figure}

We emphasize here that for these results, relative group size matters. An example of such group size effect can be seen in Figure \ref{fig:BB_size}. Here, we see that both stabilizing, as well as destabilizing effects, depend on the relative group size and the type of population heterogeneity. Figure \ref{fig:BB_size} (a) shows only for intermediate relative group size can heterogeneity in vaccine coverage induce a backward bifurcation. In contrast, Figure \ref{fig:BB_size} (b) shows how heterogeneity in vaccine efficacy cannot cause forward bifurcations for intermediate relative group size.

\subsection{Impact of homophily} \label{subs:hom}

Finally, we want to study the effect of homophily on equilibrium disease burden in our model. To this end, we resort to numerical simulations instead of analytical solutions. 

We want to understand how the dynamical behavior of the model changes in the presence of homophily, that is, individuals prefer to interact with similar others. Thus we do not necessarily have $C_{i,i} = C_{i,j} = 1$ but rather $C_{i,i} > 1 > C_{i,j}$. For the sake of comparison and also to keep the effective $R_0$ constant, we assume that $C_{1,1} + C_{1,2} =C_{2,2} + C_{2,1} = 2$. It has been found by \cite{vc-hom1} as well as \cite{vc-hom2} that the presence of homophily can either decrease or increase the proportion of people that get infected in SIR models. In this work,  we focus on how the occurrence of endemic equilibria depends on the level of homophily instead.  We begin by looking at the basic reproductive ratio $R_0$, which here is given by 
\begin{align*}
    R_0&= \frac{1}{2(\gamma + \mu)} C_{1,1}\left( \beta_1 P_1 \frac{\psi+\mu+(1-\varepsilon_1)\phi_1}{\psi+\mu+\phi_1} + \beta_2 P_2 \frac{\psi+\mu+(1-\varepsilon_2)\phi_2}{\psi+\mu+\phi_2} \right. +\\
    &\left. \left(\left( \beta_1 P_1\frac{\psi+\mu+(1-\varepsilon_1)\phi_1}{\psi+\mu+\phi_1} + \beta_2 P_2\frac{\psi+\mu+(1-\varepsilon_2)\phi_2}{\psi+\mu+\phi_2}  \right)^2 \right.\right.\\
    &\left.\left.- 4\beta_1 \beta_2 P_1 P_2 \frac{\psi+\mu+(1-\varepsilon_1)\phi_1}{\psi+\mu+\phi_1}\frac{\psi+\mu+(1-\varepsilon_2)\phi_2}{\psi+\mu+\phi_2}\left(1-\frac{C_{1,2}^2}{C_{1,1}^2} \right)\right)^{1/2}\right).
\end{align*}
Hence, $R_0$ is increasing in $C_{1,1}$, i.\,e.\ contact within groups, and homophily might significantly impact whether a disease can spread within a population and become endemic. This can be intuitively explained as follows. Typically we have $R_0>1$ in one of the groups, while $R_0<1$ in the other group. Thus in the fully well-mixing scenario, the latter group prevents an outbreak from happening. As people interact more with people within their group, the disease can break out and persist within the group with larger $R_0$, also affecting the group that has smaller $R_0$ because of their intergroup interactions.

\begin{figure}[tb]
    \centering
    \includegraphics[width=\linewidth]{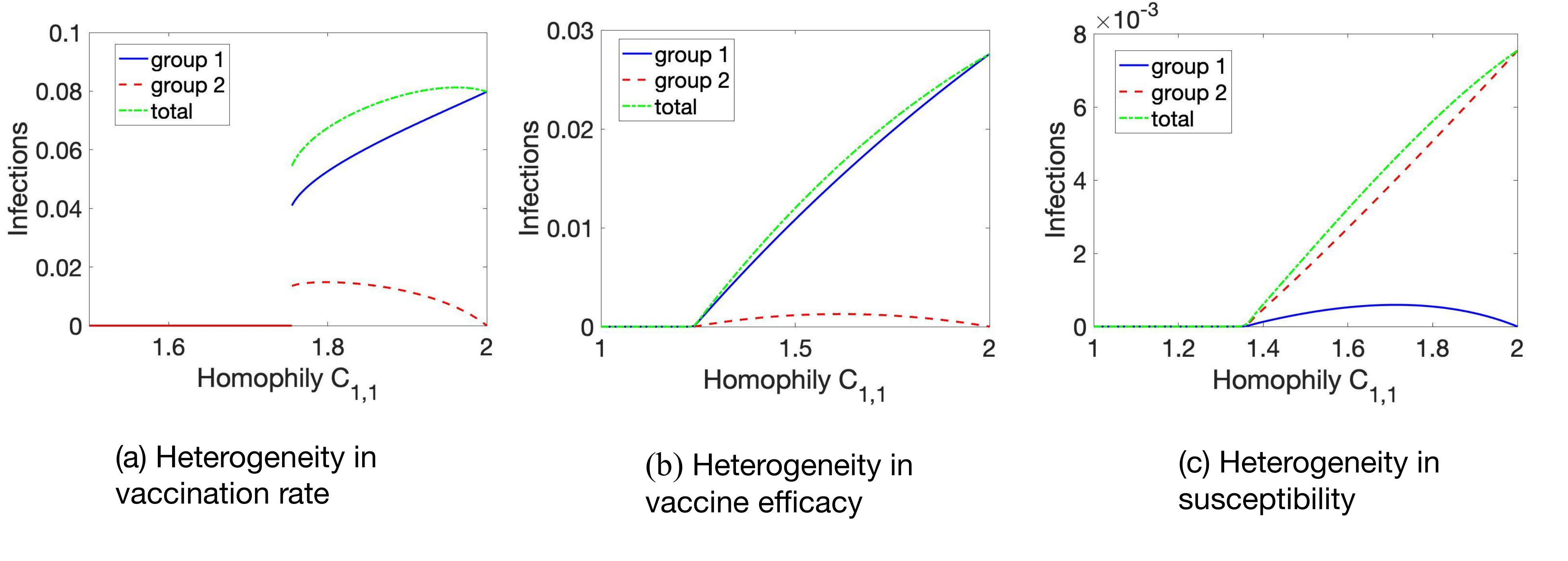}
    \caption{Homophily in heterogeneous populations can cause the emergence of endemic equilibria. We observe a backward bifurcation when increasing homophily in the presence of heterogeneity in vaccine coverage in (a) as compared to a forward bifurcation in the presence of heterogeneity in vaccine efficacy in (b) and in transmission rate in (c). Parameters are Effect of (a) $\beta = 0.14,\gamma = 0.05,\varepsilon = 0.7,\nu = 1/30,\psi = 1/(5*365),\phi = 0.105,\delta_\phi =0.095$, (b) $\beta = 0.18,\gamma = 0.1,\varepsilon = 0.55,\nu = 1/50,\psi = 1/100,\phi = 0.1,\delta_\varepsilon =0.35$, and (c) $\beta = 0.125,\gamma = 0.05,\varepsilon = 0.65,\nu = 1/300,\psi = 1/300,\phi = 0.1,\delta_\beta =0.05$. We use the Matlab method ode23, which is an implementation of the Bogacki–Shampine method—an explicit Runge–Kutta (2,3) pair, to get these results. }
    \label{fig:homophily}
\end{figure}

This observation applies to all types of heterogeneity we investigated.  For a fully well-mixing population, a small number of people that are infected do not lead to an outbreak, and the only equilibrium is the disease-free equilibrium. As people prefer to interact with people in the same group more, within group $R_0$ passes the threshold $R_0=1$ and the disease might become endemic. Note that we might observe both, a forward or backward bifurcation (see Figure \ref{fig:homophily}).  Interestingly, the chosen model parameters can cause a backward bifurcation in the corresponding well-mixed, homogeneous population. This backward bifurcation is preserved by the presence of heterogeneity in vaccine uptake (Figure \ref{fig:homophily}a). However, heterogeneity in vaccine efficacy (Figure \ref{fig:homophily}b) or transmission rate (Figure \ref{fig:homophily}c) introduces a forward bifurcation instead. 

\section{Discussion \& Conclusion}
\label{sec:disc}

It has been shown that the presence of heterogeneity in a population of susceptibles, e.\,g.\ through vaccination can cause the emergence of backward bifurcations (see e.\,g. \cite{HADELER1995,HADELER1997,Brauer2004,KRIBSZALETA2000,HUI2005,KRIBSZALETA2002,NUDEE2019,GARBA2008,bb-ex,VANDENDRIESSCHE2002,GUMEL2012}). However, the effect by introducing further heterogeneity among groups on this backward bifurcation still has to be understood fully. In this work, we examine the effect of additional heterogeneity on the spread and control of infectious disease dynamics by introducing groups with different vaccine coverage, vaccine efficacy, and susceptibility. We find an explicit formula for the equilibria of the simplified two-group SIRV model. Particularly, heterogeneity in vaccine coverage can greatly induce endemic equilibria and backward bifurcations. Heterogeneity in vaccine efficacy and susceptibility each can have additional subtle effects on this dynamical behavior.

Another common assumption in some prior models is well-mixing. This work stresses how substantially homophily can affect the dynamics of endemic diseases. When we relax the well-mixing assumption and assume that people prefer to interact with others from the same group, the resulting $R_0$ increases and might cause the disease to become endemic. \cite{salathe2008effect,hom-het1,hom-het2,vc-rates,vc-rate-het} have found that people that are skeptical of vaccination appear to prefer interacting with each other. In this work, we find that homophily in groups with different vaccine coverage (through different vaccine uptake rates) can cause a disease to become endemic along with backward bifurcation in the presence of imperfect vaccines. On the other hand, the presence of homophily for groups with different vaccine efficacies or susceptibilities appears not to exhibit the backward bifurcation behavior for the same model parameters considered, while still causing the disease to become endemic. This result emphasizes that homophily is an important factor in models for causing disease endemicity that should be further investigated.

To summarize, we analyze the joint effect of population heterogeneity and homophily in an endemic disease model with two groups. Interestingly, heterogeneity in vaccine coverage can induce backward bifurcations, while the presence of homophily has a profound effect on the bifurcation dynamics and can also facilitate the emergence of endemic equilibria. As the world embraces COVID-19 as an endemic disease, mass vaccination remains a major intervention to manage the disease. However, vaccination coverage is hugely heterogeneous across nations and regions and even across local communities.
In light of this, our modeling results  emphasize the importance of population heterogeneity and homophily and will have practical implications in the spread and control of infectious diseases in the post-pandemic era.

\enlargethispage{20pt}

\section*{Competing interests}
We have no competing interests to declare.

\section*{Acknowledgement}
F.F. is grateful for support from the Bill \& Melinda Gates Foundation (award no. OPP1217336), the NIH COBRE Program (grant no.1P20GM130454), the Neukom CompX Faculty Grant, the Dartmouth Faculty Startup Fund, and the Walter \& Constance Burke Research Initiation Award.

\section*{}


\end{document}